\documentclass[journal]{IEEEtran} 				

\IEEEoverridecommandlockouts 						
\usepackage{amsmath,amssymb}
\usepackage{amsthm}
\usepackage{color}
\usepackage[mathscr]{eucal}
\usepackage{graphics,graphicx,multicol}
\usepackage{epsfig}
\usepackage{enumerate}
\usepackage{subfigure}
\usepackage{algorithmic}
\usepackage{algorithm}
\usepackage{morefloats}
\usepackage{enumerate}
\usepackage{multirow}
\usepackage{array}
\usepackage{pstricks, pst-node, pst-plot, pst-circ}
\usepackage{moredefs}
\usepackage{courier}
\usepackage[style=ieee,backend=bibtex]{biblatex}
\bibliography{mybibliography.bbl}
\usepackage{url}

\begin{document}

\title{\vspace{-2ex} Secure Authentication of ADS-B Aircraft Communications using Retroactive Key Publication}
\author{Pavana Prakash, Ahmed Abdelhadi, \IEEEmembership{Senior Member,~IEEE},  Miao Pan, \IEEEmembership{Senior Member,~IEEE}
\vspace{-4ex}

\thanks{The authors are with University of Houston, Houston, TX, 77004, USA. Emails: pprakash3@uh.edu, aabdelhadi@uh.edu, mpan2@uh.edu.}}

\maketitle

\begin{abstract}
Automatic Dependent Surveillance-Broadcast(ADS-B), is the next generation Air Traffic management system to monitor the airspace for air traffic communication and traffic information. While the ADS-B empowers aircrafts to broadcast their location information automatically and provide situational awareness, it is susceptible to attacks and security issues. In this paper, we introduce a method to secure the ADS-B protocol in aircraft communication using Retroactive Key Publication where senders publish their keys retroactively, which is different from the traditional asymmetric cryptography. The deduced solution does not rely on a connection or two-way packets exchange to establish security. It compensates for the loss of packets owing to huge air traffic, yet preserving the open and broadcast nature of ADS-B. Our proposed protocol uses the existing ADS-B system and same hardware with no modifications but still adds security. Our secure system has low impact on current operations and retains the operational efficiency of the current aircraft system.
\end{abstract}


\pagenumbering{gobble}
\vspace{-3ex}

\section{Introduction}\label{sec:intro}

Air Traffic Organization (ATO) of the Federal Aviation Administration (FAA) provides service to more than 43,000 flights and 2.6 million airline passengers across more than 29 million square miles of airspace everyday\cite{FAA}. With the growing adoption of unmanned areal vehicle (UAV) technology for civil applications, a further boost in air traffic may be expected over the coming years\cite{Realities}. In such a dense airspace, co-ordination and operating by rules is fundamental to avoid collisions and accidents. Though the existing Mode-S is the most popular tracking system, it relies on radar rotation and transmits information about the aircraft to the SSR (Secondary Surveillance System) system only upon interrogation. Hence, it requires a two-way communication to convey a single piece of information. An emerging problem with the existing radar system i.e., traditional PSR (Primary Surveillance System) and SSR systems is their relatively low precision and detection accuracy\cite{RTCA}. Hence, a new technology to overcome the above stated issues, achieve the required precision and replace conventional radar systems, is the Automatic Dependent Surveillance Broadcast system (ADS-B).

The FAA has mandated to equip ADS-B system in all aircraft in the United States by 2020 to monitor air traffic\cite{Security}. ADS-B repurposes the old Mode-S equipment, so that the same frequency as Mode-S, 1090 MHz can be used by ADS-B as well. This move from old tracking methods to the new one will subsequently reduce the cost and increase tracking effectiveness. Aircraft equipped with an ADS-B Out transmitter send their position, altitude, heading, speed etc. to a network of ground stations that relays the information to Air Traffic Control (ATC) displays\cite{ADSB}.

While the ADS-B empowers aircrafts to broadcast their location information automatically and provide situational awareness, it is susceptible to attacks and security issues. In this paper, we introduce a method to secure the ADS-B protocol in aircraft communication using Retroactive Key Publication where, senders publish their keys retroactively. This allows a system to establish a Public Key Infrastructure (PKI) with symmetric keys. 

The deduced solution overcomes security issues in the ADS-B protocol in the Air Traffic Management System and prevents cyber-attacks like message forgery and modification attacks. Packet interception can be detected easily and such packets are dropped. It preserves the broadcast nature of the system and does not rely on multiple packets or a connection to establish security. Since aircraft travel at high speeds and communicate over long distances, packet loss is inevitable. Hence, along with counteracting the above stated facets, our security measure also tolerates loss of packets and is robust to information loss.

The remainder of this paper is organized as follows.
Section \ref{sec:System_model} provides the details of the system model being used. Section \ref{sec:Proposed Protocol} proposes the use of Retroactive Key publication concepts to implement source authentication for ADS-B. In Section \ref{sec:Algorithm}, we present our Authentication Algorithm for the key generation and packet transmission. Section \ref{sec:simulation} discusses simulation setup and provides quantitative results along with discussion. Section \ref{sec:conclusion} concludes the paper.


\begin{figure}[h]
\centering
\includegraphics[width=88mm]{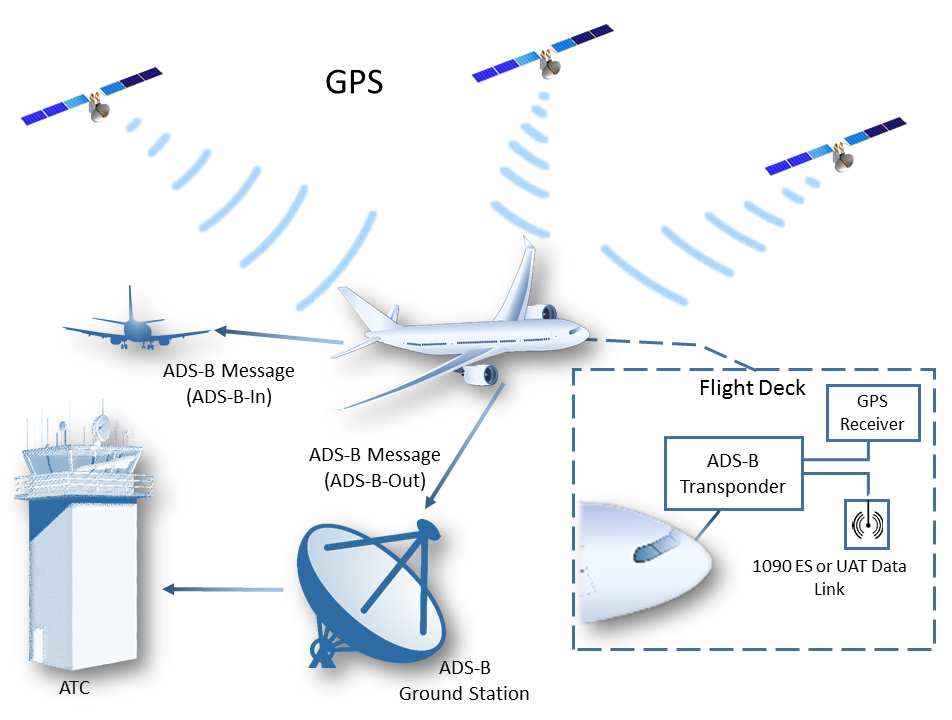}
\caption{Aircraft communication using ADS-B}
\label{fig:fig1}
\vspace{-2.5ex}
\end{figure}
\vspace{0ex}

\section{System Model}\label{sec:System_model}
ADS-B, the satellite-based successor of radar as stated by the FAA, is the Next Generation (NextGen) air traffic management system to monitor the airspace for air traffic communication and traffic information. ADS-B improves situation awareness of pilots by even including weather reports and temporary flight restrictions, thereby providing better visibility in harsh conditions too. In order to use ADS-B, every aircraft needs to be fitted with ADS-B OUT to broadcast information. Anyone with ADS-B IN, will receive this information. Since the FAA has only mandated that aircrafts have ADS-B OUT, not all aircrafts have ADS-B IN. The ATC, however, must have ADS-B IN in order to receive communications from aircrafts to keep track of local traffic.

The Aircraft communication using ADS-B is shown in Figure \ref{fig:fig1}. In ADS-B, each aircraft determines its position and velocity information from the GPS data using an onboard GPS receiver. This information is periodically broadcasted in a message by the transmitting subsystem, ADS-B OUT. Most information provided by ADS-B is broadcasted periodically (e.g., the position twice per second) while the transmission of other types (e.g., status or intent) is event-driven\cite{Security}. The location and velocity are each broadcasted in separate packets twice a second by the aircraft. The call sign is sent once in every five seconds and other information such as aircraft intent, identification, urgencies and uncertainty level are sent at lower intervals which account to a total of two packets per second. Hence, leading to an average rate of 4.2 messages per second. With the current standard recommended, this could lead up to a maximum transmission rate of 6.2 messages per second\cite{RTCA}.

ADS-B uses Pulse Position Modulation (PPM) to encode information since it is relatively robust against interference and collisions. Transmission of a bit lasts for one microsecond. The system uses 1090 MHz as the carrier frequency in order to maintain compatibility with legacy systems. Since the previously mandated Mode-S transponders operate at 1090 MHz frequency for downlink (air-to-ground) communication, fitting an ADS-B OUT transponder will be a simple upgrade for monetary reasons too.

These messages are then received by the ADS-B Ground Stations and other nearby aircraft equipped with the receiving subsystem, ADS-B IN. The aircraft's position, trajectories and other information is transmitted to the ATC. The controllers on the ground also communicate this information to one another as aircraft transition between their areas of control. Unlike legacy systems with interrogation, since ADS-B broadcasts periodically, there is only one packet for exchange of information and ATC just has to keep listening. 

\begin{figure}[h]
\centering
\includegraphics[width=0.48\textwidth]{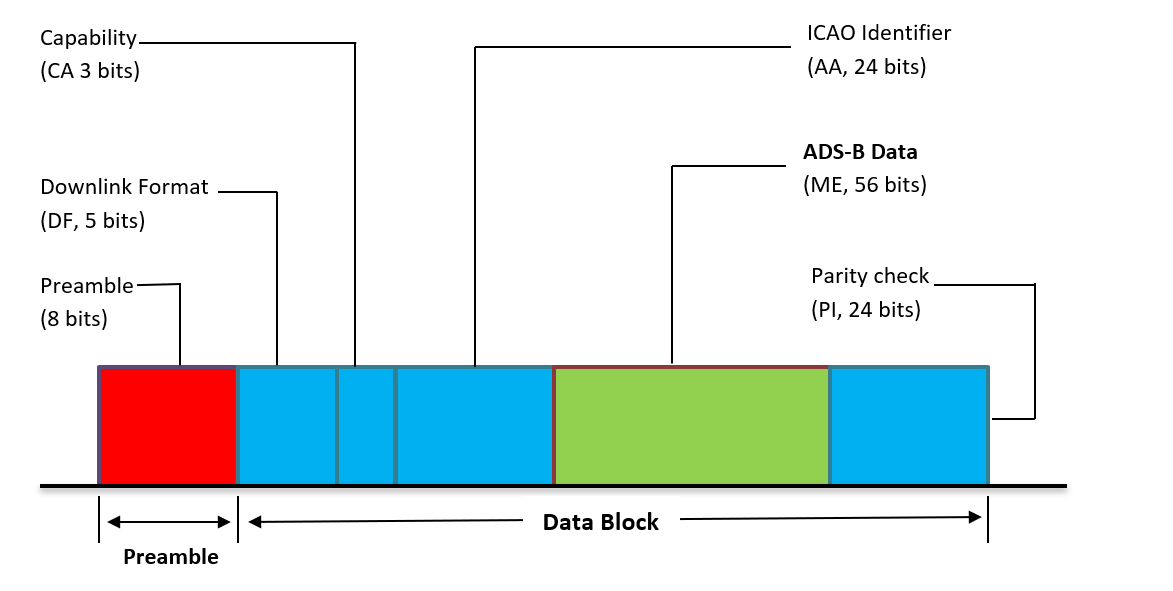}
\caption{ADS-B packet format}
\label{fig:fig2}
\vspace{-1ex}
\end{figure}

Figure \ref{fig:fig2} shows the format of an ADS-B packet. The aircraft broadcasts an 8 microsecond preamble before the packet to allow synchronization on reception. The ADS-B packet begins with the 5-bit Downlink Format (DF) which indicates the message type and has a value of 0x17 for ADS-B Extended Squitter. The next 3 bits is a subtype and additional identifier describing the capabilities the participant possesses. The 24-bit ICAO Identifier is a unique alphanumeric code assigned to each aircraft transponder by the International Civil Aviation Organization(ICAO). The ADS-B data is fit in 56 bits out of which, the first 5 bits indicate the Type Code that helps in identifying what information is contained in the message. ADS-B uses a cyclic redundancy check to validate the correctness of the received message, where the last 24 bits are the parity bits\cite{Sun}. The receiver can correct up to 5 bits using fixed generator polynomial.
\vspace{-0.05ex}

\section{Proposed Protocol}\label{sec:Proposed Protocol}

The Retroactive Key Publication concept implements source authentication for ADS-B. It also aims to prevent message forgery and message modification attacks. It uses the message digest SHA-256 along with hash-based Message Authentication Codes (MACs). Given the 56-bit data limitation, the system sends the original message and its correspondingly computed MAC separately instead of appending the MAC to the message and sending as one packet. Hence, each transmitting node sends pairs of packets. The first packet contains the aircraft data and the second packet contains the MAC to the first packet. To account for the potential packet loss given radio transmission through air, the node sends this pair twice. The computed key to be shared is 50 bits long and is also sent twice about after every ten packets (for possible loss). This makes the valid duration for each key as ten packet duration long.

To differentiate between the kinds of packets sent, the five bit Type Code mentioned in the previous section is used that specifies whether the packet contains a MAC or a key. So, the leftmost 50 bits of the 256-bit message digest generated by SHA-256 are selected as the truncated message\cite{Dangnist}. Strength of the algorithm after truncation would be, \(\frac{50}{2}\) = 25 bits and an expected preimage resistance of 50 bits\cite{Dangnist}. An attacker then has a \(\frac{1}{2^{5 0}}\) chance of computationally finding a value similar to our input using random bits in SHA-256.

Furthermore, to prove the integrity of the message over untrusted network, once the sender publishes the key, the recipients can perform the same HMAC algorithm on the previously buffered messages and compare the generated MAC with the MAC received and authenticate the sender. To add a further layer of security, we make sure that the key came from a legitimate source since an attacker could have fabricated a key as well. To check this, the receiver runs a pseudo-random one way function which generates keys from a seed value that is random. Also, using the SHA-256 function, the system could generate new keys through successive hashes of each key. 

\begin{figure}[ht]
\centering
\includegraphics[width=0.48\textwidth]{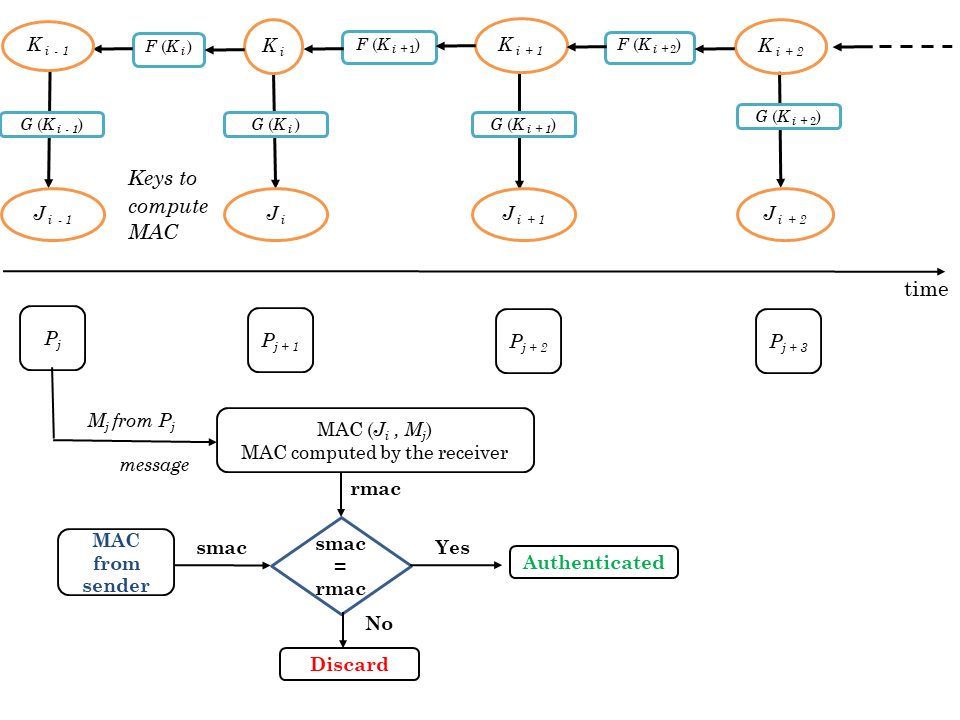}
\vspace{-1.5ex}
\caption{Our proposed authentication flow diagram}
\label{fig:fig3}
\vspace{-2.5ex}
\end{figure}

As seen in Figure \ref{fig:fig3}, the first one-way function $F$ generates the chain and the second one-way function $G$ derives the MAC keys from a seed value\cite{Security}. To avoid cryptographic weakness by using the same key for deriving the next keys as well as computing MAC, we use the second pseudo-random function $G$.   The ADS-B data along with the computed MAC is sent through every other packet $P_j$. The key to decode these packets is sent after a delay $d$, after ten packets are transmitted, through the packet $P_j$. Hence, $P_j$ could refer to packet containing flight data $M_j$, computed MAC $MAC(J_i, M_j)$ or the key $K_i$, which could look like,
\begin{equation}
    \vspace{-0.4ex}
    P_j = {M_j , MAC(J_i, M_j) , K_{i-d}}
    \vspace{-0.4ex}
\end{equation}

Receiver has to authenticate the key $K_i$ against previously received keys to ensure they are from the same key chain. This implies that for packet $P_{j+1}$, the MAC is computed based on the key $K_i$ of the corresponding time. Hence, last generated key is used first and worked back towards the seed value as time progresses. If one key gets lost in transmission, the receiver can use this one way function to regenerate that key with the next key. The key remains secret for the next $d-1$, so for example, messages sent at $j$ effectively disclose the key $K_{j-d}$\cite{Tesla}. This ensures that the data with key $K_i$ could only have been sent before the key was published i.e., before the interval $i+d$ where $d$ is the key disclosure delay.

On receiving the packet $P_j$, the receiver uses the key $K_{i-d}$ disclosed in $P_j$ to determine when the packet arrived. It then checks if the number of packets after which the key arrives is reached. So, if $i < l+d$, where $l$ is an arbitrary number of packets sent, then the packet is safe and key disclosure by the sender has not yet happened that will verify the packet $P_j$\cite{Tesla}. Hence, at this point, the receiver puts the received message or MAC \((M_j, MAC(J_i, M_j))\) into a buffer and waits to learn the key through the packet to authenticate it. On receiving the disclosed key $K_i$, the receiver checks if this is the latest key received to date. Then, to check its legitimacy, it verifies an earlier key; for example, $K_{i-v} (v<i)$ that \(K_{i-v} = F^v(K_i)\)\cite{Tesla}. Then runs the one-way function $G(K_i)$ to compute the key $J_i$ for HMAC computation. If this computed MAC is the same as the MAC received through the packet from the sender, the message is deemed valid.

To read more about the HMAC, refer to the HMAC construction by Bellare et al.\cite{HMAC}.

Variables and acronyms considered in the Authentication Algorithm are, \textit{M} Message, \textit{l} arbitrary number of packets sent, \textit{K} Key, \textit{i} Key index, \textit{d} Key disclosure delay valued 10 packets, \textit{H} Hash function - SHA-256, \textit{K'} Block-sized key after hashing down the key size, $\bigoplus$ bitwise exclusive or (XOR), $\|$ concatenation, \textit{smac} MAC computed by the sender, \textit{rmac} MAC computed by the receiver, \textit{o\_pad} Outer padding valued 0x5c, \textit{i\_pad} Inner padding valued 0x36, \textit{$P_j$} Packet at $j$th index - this could contain the ADS-B message/smac/key, \textit{$J_i$} Key to compute MAC determined using $K$, \textit{F} Pseudo-random one-way function generating the key chain, \textit{G} Pseudo-random one-way function deriving MAC from the seed value, \textit{v} An earlier key index

\vspace{-1ex}
\section{Authentication Algorithm}\label{sec:Algorithm}
\vspace{-1ex}
\begin{algorithm}
\caption{Authentication Algorithm}\label{alg:Algo}
\begin{algorithmic}[1]

\STATE {smac = $H(K_i, M_j) = 
\newline H((K_i' \bigoplus \text{o\_pad}) \| H(K_i' \bigoplus \text{i\_pad}) \| M_j)$}

\WHILE{$P_j$ is received,} 
    \STATE $l = 0$
    \newline Wait until the packet count is reached to disclose the key
    \WHILE{$i$ less than or equal to \textit{(l + d)},}
        \STATE buffer := $(i, M_j, \text{smac})$
    \ENDWHILE
    
    Now set the arbitrary counter to the latest packet count
    \STATE $l = i
    $
    \newline Check if $K_i$ is the latest key received by using it to generate the previous keys
    \IF{$K_{i-v} == F^v(K_i)$}
        \STATE $K_i = F(K_i)$
        \STATE Extract key to compute MAC
        \STATE $J_i = G(K_i)$
        \STATE{rmac = $H(J_i, M_j) = 
        \newline H((J_i' \bigoplus \text{o\_pad}) \| H(J_i' \bigoplus \text{i\_pad}) \| M_j)$}
        \IF{rmac == smac} 
            \STATE {Valid message} 
        \ELSE 
            \STATE{Invalid message} 
        \ENDIF
    \ELSE
        \STATE {Key is invalid}
    \ENDIF
\ENDWHILE
\end{algorithmic}
\end{algorithm}
\vspace{-3ex}
\section{Simulation Results}\label{sec:simulation}

This section simulates the performance of the new ADS-B system also including the previously widely used Mode-S. Since transmission by these transponders occur continuously and independently of the other, this can be considered as a stochastic process and the system can thus be modeled a Poisson process. Taking into account that a collision occurs in the time interval of $t_p$ length which represents the length of the packet, $n$ being the number of transmitters and $T$ representing the frequency of transmission for a packet type per aircraft, the probability of collisions can be calculated using the following equation from \cite{Rajba},

\vspace{-2ex}
\begin{equation}
    \vspace{-1ex}
    P(collision) = 1 - e^{(-n^\frac{t_p}{T})} - n\frac{t_p}{T}e^{(-n^\frac{t_p}{T})}
\end{equation}

Here, the first term represents the scenario when no sources send a packet and the second that only one source sends a packet during the time it takes to send one packet. Since both ADS-B and Mode-S share the same frequency but different packet length, a modified version of the above equation helps us model the collision probability when both the transponders are transmitting. Using the values of 112 and 120 bits packet length for Mode-S and ADS-B respectively and 6.2 packets per second frequency in equation \ref{eqn:eqn3} below, Figure \ref{fig:fig4} is obtained.

\vspace{-2ex}
\begin{equation}
\label{eqn:eqn3}
    P(collision) = 1 - P_A(0)P_S(0) - P_A(0)P_S(1) - P_A(1)P_S(0)
\end{equation}

where, $P_A(0)P_S(0)$ indicates no sources send ADS-B or Mode-S packets,
$P_A(0)P_S(1)$ implies that sources send only one Mode-S packet and $P_A(1)P_S(0)$ means that sources send only one ADS-B packet. 

\vspace{-3ex}
\begin{figure}[h]
\centering
\includegraphics[width=0.48\textwidth]{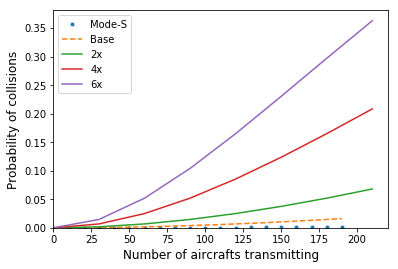}
\vspace{-1ex}
\caption{Probability of Packet Collisions}
\label{fig:fig4}
\vspace{-1ex}
\end{figure}

Figure \ref{fig:fig4} shows the probability of collisions between packets on the 1090 MHz frequency, which includes both Mode-S and ADS-B. The graph assumes the full 6.2 packets per second transmission. The result contains the scenario where only Mode-S is transmitting, then with both Mode-S and ADS-B transmitting on the same frequency, 1090 MHz at base transmission rate. The plot then displays this with increasing transmission rates to show the probability of collisions. We see that, even with our security measure added, the existing performance is not affected much.

\vspace{-1.75ex}
\section{Conclusion}\label{sec:conclusion}
In this paper, we introduced a method to secure the NextGen air traffic monitoring technology, ADS-B. Our new method preserves the broadcast nature of the system and does not rely on two-way communication or a connection to establish security. Yet, it has low impact on the the actual performance of the system and does not affect the current operations. It overcomes security issues in the ADS-B protocol in the Air Traffic Management System and prevents cyber-attacks like message forgery and modification attacks. Packet interception can be detected easily and such packets are dropped. Along with counteracting the above stated factors, our security measure also tolerates loss of packets and is robust to information loss. All of these, without any changes to the existing system or a hardware upgrade.

\nocite{EUROCONTROL}
\nocite{ModeS}
\nocite{changStory}
\nocite{Opensky}
\nocite{Sos}
\printbibliography

\end{document}